\documentclass{raa}  

\usepackage{graphicx,times}
\usepackage{natbib}
\usepackage{amssymb,amsmath}
\bibpunct{(}{)}{;}{a}{}{,}

\usepackage[pagebackref=true]{hyperref}

\begin{document}

	\title{Unveiling the nature of polar-ring galaxies from deep imaging}

 \volnopage{ {\bf 20XX} Vol.\ {\bf X} No. {\bf XX}, 000--000}
   \setcounter{page}{1}

   \author{Aleksandr V. Mosenkov
      \inst{1, 2}
   \and Vladimir P. Reshetnikov
      \inst{2, 3}
   \and Maria N. Skryabina
      \inst{3}
   \and Zacory Shakespear
      \inst{1}   
   }

   \institute{Department of Physics and Astronomy, N283 ESC, Brigham Young University, Provo, UT 84602, USA; {\it aleksandr\_mosenkov@byu.edu}\\
     \and
     Central (Pulkovo) Astronomical Observatory, Russian Academy of Sciences, Pulkovskoye Chaussee 65/1, 196140 St.Petersburg, Russia\\
     \and
     St.Petersburg State University, 7/9 Universitetskaya nab., 199034 St.Petersburg, Russia\\
\vs\no
   {\small Received~~20xx month day; accepted~~20xx~~month day}}

\abstract{General structural properties and low surface brightness tidal features hold important clues to the formation of galaxies.
In this paper, we study a sample of polar-ring galaxies (PRGs) based on optical imaging from the Sloan Digital Sky Survey (SDSS) Stripe\,82 and other deep surveys. We investigate the deepest images of candidates for PRGs to date. We carry out photometric decomposition on the host galaxies and associated polar structures that allows us to derive the structural properties of both components. We are able to detect very faint tidal structures around most PRGs in our sample. For several galaxies, we can directly observe the formation of the polar ring due to merging, which is manifested in debris of the victim galaxy and an arc-like polar structure made up of its material. In a few cases, we can discern signs of tidal accretion. The results obtained indicate that the gravitational interaction and merging of galaxies are the most plausible mechanisms for the formation of polar-ring galaxies.
\keywords{methods: data analysis --- techniques: photometric --- galaxies: structure}
}

   \authorrunning{A. Mosenkov et al. }            
   \titlerunning{Deep imaging of PRGs}  

   \maketitle

%
%
\section{Introduction}           
\label{sect:intro}
Polar-ring galaxies (PRGs) are among the most beautiful and unusual objects 
in the Universe. We are used to the fact that galaxies are flat, held by 
regular rotation, or almost spherical, supported by random motions of stars.
However, there are objects that do not fit into this simple scheme.
They embody a strange symbiosis -- an early-type galaxy and a spiral one in 
the same object. Numerous examples of such objects are given in
\cite{whit1990} (= PRC, Polar Ring Catalogue), \cite{moiseev2011} (= SPRC, SDSS-based Polar Ring Catalogue) and 
\cite{rm2019} (new candidates for polar-ring galaxies).

The main characteristic feature of PRGs is that they contain two large-scale, 
morphologically and kinematically decoupled subsystems --
the central galaxy (the host) and the polar structure (PS). 
(In this paper we will use the term ``PRG'' to designate the whole
class of galaxies with near-polar optical structures, without
dividing into polar-disc or polar-ring galaxies -- e.g. \citealt{iodice2014}).
These two subsystems of PRGs show very different characteristics.
In most cases their central objects look like early-type galaxies 
without gas, with relatively red colour indices
and rotation around their minor axes (e.g. \citealt{whit1987};
\citealt{fink2012}; \citealt{rc2015}). On the other hand, PSs 
(in most cases their major axes are almost orthogonal to the major axes 
of the central galaxies) contain stars and gas, which rotate around the 
major axes of the central galaxies. They have blue colours and 
show signs of ongoing star formation; the gas in 
such structures often has sub-solar metallicity (e.g. \citealt{whit1987};
\citealt{vg1987}; \citealt{rc1994,rc2015}; \citealt{iodice2002}; 
\citealt{em2019}). The unique morphology of PRGs makes them very useful laboratories 
for studying a number of extragalactic astronomy issues linked to the formation and 
evolution of galaxies, and for exploring the properties of their dark haloes
(see \citealt{khop2014}, \citealt{comb2014} and references therein).

To explain the observed structure of PRGs, a number of sometime exotic formation mechanisms were 
proposed. For instance, \citet{sersic1967} 
suggested that the PS of NGC\,4650A is a string
of clouds of relativistic electrons ejected from the nucleus of the
galaxy. For NGC\,2685 -- the first known PRG \citep{bb1959} --
it has been suggested that the central object in the galaxy represents the 
result of an explosion in a preexisting spiral galaxy outlined by the 
rings \citep{gk1979}. The structure of UGC\,7576 has been interpreted as
an edge-on spiral galaxy with a prominent bulge aligned orthogonally to
the disc \citep{mould1982}.

The most popular explanation for the structure of PRGs originates from 
Alar Toomre's comment that in the case of NGC\,2685 we may be seeing
the accretion of an intergalactic gas cloud or ``a small, gas-rich 
companion in the not too distant past'' \citep{toomre1977}. 
In the same year, a similar idea was expressed in the work of 
\cite{knapp1977}. A few years later, \cite{shane1980} discussed the 
external accretion to explain the structure of NGC\,2685 in more detail. 
\cite{schw1983} expanded the accretion scenario and included the possibility 
of ``the accretion of material transferred between two galaxies during a close
encounter''. 

Currently, there are several scenarios for the formation of PRGs due to 
interaction of galaxies with their environment, such as:

-- Major merger: a head-on collision of two galaxies with their discs oriented
perpendicularly (\citealt{bekki1997,bekki1998}; \citealt{bourcomb2003}).
This scenario is particularly able to explain the presence of two 
ring structures in ESO\,474-G26 as a transient stage in the merging of two 
spiral galaxies \citep{rbc2005}. Also, the delayed return of tidally ejected
material during the merging can lead to the formation of PSs 
(\citealt{hib1995}).

-- Minor merger: a tidal disruption of a small companion in a near-polar orbit
of the host galaxy. This mechanism could only explain narrow unclosed 
rings (e.g. \citealt{john2001}, \citealt{rbc2006}).

-- Tidal accretion: a gas-rich donor galaxy loses matter to form the ring
around the host (\citealt{rs1997}, \citealt{bourcomb2003}). The donor
galaxy must be on a nearly polar orbit with respect to the 
host galaxy. An example of applying this scenario to a giant spiral 
galaxy AM\,1934-563 is presented in \cite{rbc2006}. 
In addition, several interacting systems are observed in which matter is 
directly transferred from one member to another with the formation of a
large-scale PS (e.g. NGC\,3808A/B, NGC\,6285/86 --
\citealt{rhy1996}, \citealt{ord2016}).

-- Infall of gas from intergalactic space: a cold accretion of gas along 
cosmological filaments (\citealt{1996ApJ...461...55T,1998ApJ...506...93T}, \citealt{maccio2006}, \citealt{brook2008}).
A candidate to a PRG formed in this way is described by
\cite{stan2009}.

As such, half of these scenarios are related to galaxy mergers which underpin our current understanding of how galaxies grow and evolve under the $\Lambda$CDM paradigm.
However, it is likely that, depending on the spatial environment and properties of 
the galaxies themselves, different mechanisms for the formation of 
PRGs can be realised. To understand how a particular object formed, 
we have to acquire detailed photometric and spectral observations, as well as carry out a 
study of its environment.

A very important window for looking into the past of any galaxy is its 
study at very low levels of surface brightness. Ultra-deep images of galaxies 
can reveal the ``scaffolding'' preserved from the time of their formation --
tidal tails and bridges, stellar haloes, envelopes, plumes etc.
If the formation of PRGs is associated with accretion of matter, 
we can expect to detect traces of these processes at the periphery of galaxies.
Unfortunately, deep images of PRGs are very rare. For instance, \cite{fink2010} 
presented a deep image of a candidate to PRGs in the Subaru Deep 
Field. They described several extended low surface brightness (LSB) features
around the galaxy which can indicate an ongoing merging with a companion.
\cite{fink2012} discussed the morphology of 16 PRGs candidates based on
the original photometry and the SDSS Stripe\,82 data \citep{jiang2008}.
They noted, in particular, signs of recent interactions in several 
galaxies (asymmetric tails, stellar fans etc.). 

In this \textit{pilot} study, we exploit ultra deep images of 13 PRG candidates from
 the SDSS Stripe\,82 and other modern deep optical surveys in order to shed more light on the possible origin 
of the PSs. We focus on revealing LSB structures in and around PRGs and pay special attention to the quantitative and qualitative description of these features. Temporal fine structures, such as tidal tails, streams and shells with survival times $\sim1$~Gyr \citep[see e.g.][and references therein]{2019A&A...632A.122M,2020A&A...640A..38R}, should be good indicators of the forming PRG structures. This allows us to carry out a detailed photometric analysis of the deep images of PRGs. We also perform photometric decomposition for the entire sample and study the properties of the host and PS. 

This paper is outlined as follows. We present our sample in Section~\ref{sec:sample}. Section~\ref{sec:data} gives a description of the data and image preparation. In Section~\ref{sec:results}, we present in detail the properties for each of the selected galaxies. We discuss our results in Section~\ref{sec:discussion}. Our findings are summarised in Section~\ref{sec:summary}. Throughout this article, we adopt a standard 
flat $\Lambda$CDM
cosmology with $\Omega_m$=0.3, $\Omega_{\Lambda}$=0.7, 
$H_0$=70 km\,s$^{-1}$\,Mpc$^{-1}$. All magnitudes and colours in the paper are given
in the AB-system and corrected for the Milky Way extinction \citep{schfin2011} and $k$-correction \citep{chil2010}.

\section{The sample}
\label{sec:sample}
\begin{table}
\bc
\begin{minipage}[]{0.86\textwidth}
\caption[]{Basic properties of the sample galaxies. We adopt the original classification from the SPRC.\label{tab:Table1}}\end{minipage}
\setlength{\tabcolsep}{1pt}
\small
  \begin{tabular}{cccccccc}
  \hline\noalign{\smallskip}
SPRC type & Galaxy & RA       & Dec.    & $z$ & $M_r$ & $g - r$ & Faint features  \\ 
                &            & (J2000)& (J2000)&        & (mag)  &             &                        \\
                &            &            &            & (1)   &   (2)     &     (3)    &        (4)            \\
  \hline\noalign{\smallskip}
Best          &SPRC-1& 00:09:12&-00:36:55    &0.07325 &  -22.01 & 0.80 & Diffuse light, tidal tails and streams,    \tabularnewline
candidates &          &               &                    &              &              &         & faint polar halo   \tabularnewline
           &SPRC-4    & 02:42:58 & -00:57:09&0.04299 &  -19.43 & 0.57 & Tidal streams    \tabularnewline
           &SPRC-69  & 20:48:06&+00:04:08&0.02467 &  -19.66 & 0.88 & ---    \tabularnewline
\hline
Good       &SPRC-73  &00:32:10&+01:08:37&0.05909 & -21.31 & 0.76 & Curved tidal tail  \tabularnewline
candidates &SPRC-74 & 00:48:12&-00:12:56 &0.05646 & -21.12 & 0.73 & Curved tidal tail, arc, plume, bridge  \tabularnewline
           &SPRC-76  & 01:21:29&+00:37:29&0.04356 & -19.56 & 0.77 & Curved tidal tail, plume, bridge \tabularnewline
           &SPRC-77  & 01:58:58 & -00:29:23&0.08112 & -20.41 & 0.81 & Oval polar envelope, bridge? \tabularnewline
           &SPRC-185 &23:12:32&-00:06:37&0.02783 & -20.45 & 0.74 & Boxy polar envelope, plume, bridge? \tabularnewline
\hline
Related    &SPRC-186 & 00:48:10 & -00:54:44 & 0.03397 &-21.17 & 0.76 & ---   \tabularnewline
objects    &SPRC-188 & 03:34:06 & +01:05:40 & 0.04748 &-21.21 & 0.81 & Tidal streams, plume   \tabularnewline
           &SPRC-234 & 21:23:39 & -00:22:35 & 0.06198 &-20.51 & 0.79 & Very faint grand-design spiral structure or\tabularnewline
           &                 &               &                  &               &            &          & (un?)closed polar ring or projection?   \tabularnewline
           &SPRC-238 & 23:39:58 & +00:08:07 & 0.06018 &-21.70 & 0.62 & Straight tidal tail, arc, bridge?   \tabularnewline
\hline
Possible   &SPRC-275 & 20:53:53 & -00:58:05 & 0.10610 &-21.01 & 0.82 & ---   \tabularnewline
face-on ring &       &        &       &        \tabularnewline
  \noalign{\smallskip}\hline
  \end{tabular}
\ec
\tablecomments{0.86\textwidth}{(1) redshift;
     (2), (3) absolute magnitude in the $r$ band and $g-r$ colour;
     (4) see Sect.~\ref{sec:results} where we describe each galaxy in detail.}
 \end{table} 

Our sample is composed of 13 objects from the SPRC (\citealt{moiseev2011}) located in the 
field of Stripe\,82. The list of galaxies, along with their general
characteristics taken from the SDSS DR15 (\citealt{aguado2019}), 
is presented in Table~\ref{tab:Table1}. The galaxies in the table are divided into 4 
groups in accordance with the SPRC original classification:
three galaxies are classified as the best candidates to PRGs, 
five are good candidates, four objects are galaxies possibly related to PRGs, and
one object is classified as a possible face-on ring. As can be seen
from Table~\ref{tab:Table1}, the sample galaxies are bright objects with $g-r$ colour 
typical of PRGs (e.g. \citealt{rm2019}).

Our visual inspection of the SDSS frames reveals that most of the
sample galaxies indeed demonstrate pronounced PSs (see Sect.~\ref{sec:data}). 
In 8 cases, the PSs look symmetric (SPRC-1, 4, 69, 73, 77, 185,
188, 234), and in three they show a large-scale asymmetry (SPRC-74, 76, 238).
The visible orientation of two galaxies (SPRC-186 and 275) and our photometric decomposition (see Sect.~\ref{sec:results}) do not allow us to conclude that they have polar features. Sect.~\ref{sec:results} provides a detailed description of each object from our sample. In Table~\ref{tab:Table1}, we briefly summarise our notes on each object.

\section{Data and Preparation}
\label{sec:data}

For our study, all images of the selected galaxies are taken from the SDSS Stripe\,82 Imaging Data reduced by \citet{2014ApJS..213...12J}. To prepare galaxy images, we employ a semi-automatic image preparation pipeline IMage ANalysis\footnote{\url{https://bitbucket.org/mosenkov/iman_new/src/master/}} \citep[IMAN, see][]{2020MNRAS.497.2039M}.
It is important to note that standard SDSS frames have a depth of 26.5~mag\,arcsec$^{-2}$ in the $r$ band, whereas the average depth of the images, downloaded from the Stripe\,82 database in the same band, is $28.89\pm0.25$~mag\,arcsec$^{-2}$. Here and below we compute the photometric depth as a $3\sigma$ fluctuation in square boxes of  $10\arcsec \times 10\arcsec$. The small standard deviation of the photometric depth for our sample shows that the presence of low-contrast structures in the images under study should be detected nearly in the same way for all galaxies. Below, we briefly describe our methodology for preparing stacked cut-outs of the sample galaxies (see \citealt{2020MNRAS.497.2039M} for detail). To exploit the Stripe\,82 data to their fullest, that is to increase the depth of these observations and unveil very faint structures around the galaxies, we stack the Stripe\,82 images in the three $gri$-bands (the $u$ and $z$ bands are shallow as compared to the $gri$ bands, \citealt{1996AJ....111.1748F}), similar to what was done by \citet{2011A&A...536A..66M} and \citet{2018A&A...614A.143M}) but for standard SDSS data.

First, we employ {\tt{SExtractor}} \citep{1996A&AS..117..393B} to create a segmentation map for each image. In this map, the size of every object is increased by a factor of 1.5 to reduce the scattered light from the detected sources when subtracting the sky background. Then, we iteratively approximate the sky using polynomials, starting with 1st order. If necessary, we increase the polynomial order for a less flat sky background. Fortunately, all images do not show a significant non-flat background or Galactic cirri (only SPRC-1 and SPRC-188 have adjacent filamentary clouds with $\langle \mu_{r} \rangle\sim26.8$~mag\,arcsec$^{-2}$ in the $r$ band which do not directly overlap the galaxies). We also examined the  \citealt{1998ApJ...500..525S} reddening map and found moderate E(B-V)=$0.05\pm0.03$ for the coordinates of the sample galaxies. Therefore, we can be confident that neither the subtraction of the sky, nor dusty filaments in our Galaxy affect the LSB regions around the sample galaxies.

We carry out the above described procedures for all three $gri$ bands of the Stripe\,82 data and resample the prepared images in the $g$ and $i$ bands to the $r$ band using the {\tt{pypher}}\footnote{\url{https://github.com/aboucaud/pypher}} package \citep{2016A&A...596A..63B}. The Point Spread Function (PSF) images for this resampling are taken from the IAC Stripe 82 Legacy Project database \citep{2016MNRAS.456.1359F}. To create an extended PSF for each image, we merge this downloaded image and use it as a core PSF (seeing), with the wings (beyond 9 pixels from the PSF centre) of an unified extended PSF from \citet{2020MNRAS.491.5317I}. Finally, we stack the prepared galaxy images in the $gri$ bands using the {\tt{IRAF}}/{\tt{IMCOMBINE}} procedure. 

Applying {\tt{SExtractor}} to each of the stacked images, we select good unsaturated stars with the highest signal-to-noise ratio.  After that, we perform photometric calibration using the measured fluxes of these stars and their corresponding total magnitudes in the $r$ band from the SDSS database. By doing so, we are able to estimate the depth of the stacked images in the $r$ band. It appeared to be $29.74\pm0.13$~mag\,arcsec$^{-2}$ for the whole sample, almost one magnitude deeper than in the original deep $r$-band frames.

Finally, we mask out all sources, which do not belong to the target galaxies, using  {\tt{SExtractor}}  and the mto library\footnote{\url{ https://github.com/CarolineHaigh/mtobjects}} \citep{teeninga2015improved}. The latter package allows us to accurately mask foreground objects inside the galaxy outermost isophotes where {\tt{SExtractor}} usually faces issues with creating a correct segmentation map.

We also prepare enhanced $grz$ images (see Appendix~\ref{Appendix:supplementary}) from the DESI Legacy Imaging Surveys DR8 (hereafter DESI Legacy, \citealt{2019AJ....157..168D}), to ensure that the very faint features around the galaxies in the Stripe\,82 images are also seen in other, independently obtained images.

\section{The detailed description of the sample PRGs}
\label{sec:results}

\begin{figure*}
\label{fig:stacked_images}
\centering
\includegraphics[width=14cm]{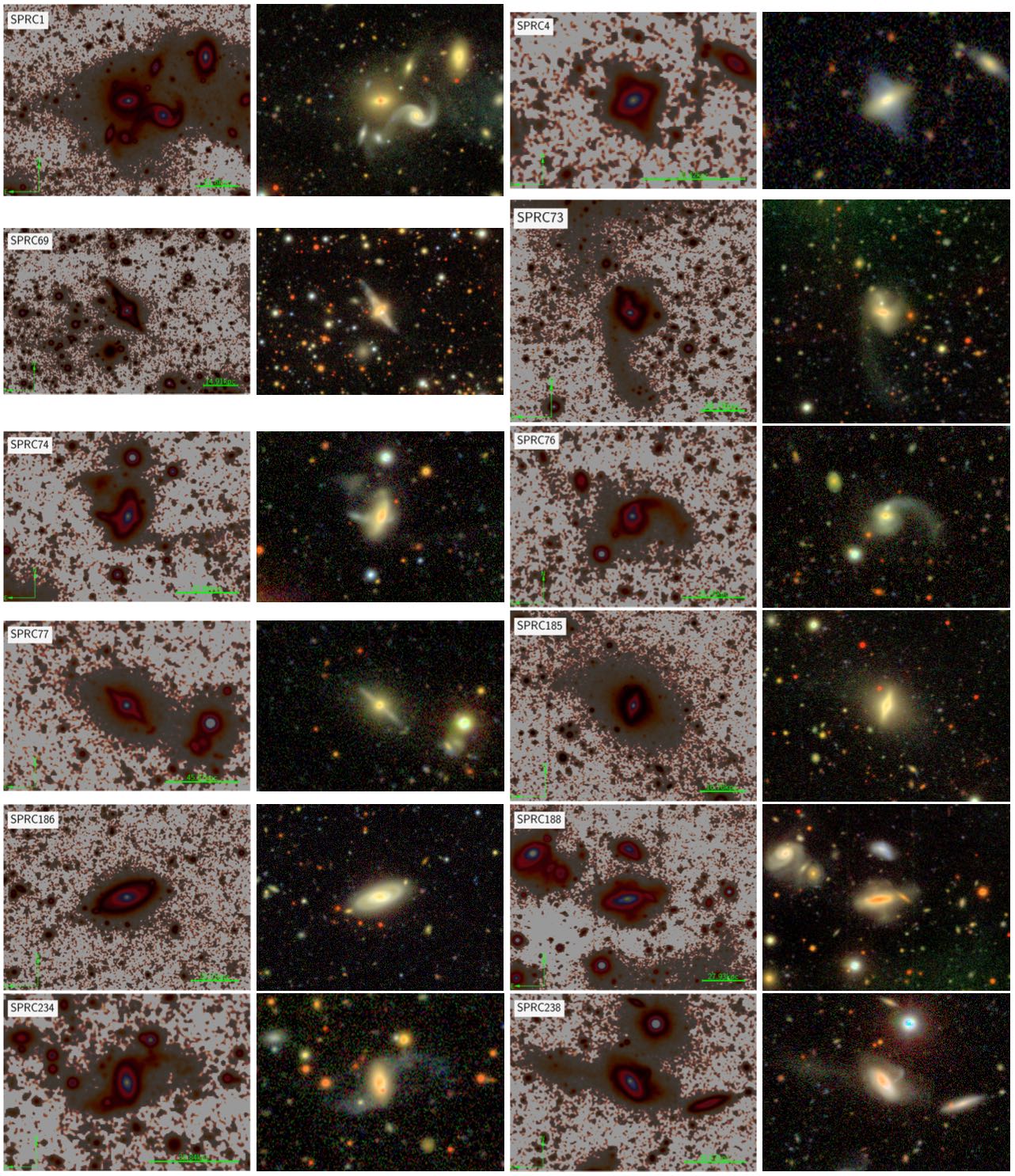}
\caption{Stacked images (left panel) and colour RGB images (right panel), created from the Stripe\,82 images in the $g$, $r$ and $i$ bands. The stacked images have been blurred with $\sigma=1$~pixel. Only pixels with surface brightnesses down to 28~mag\,arcsec$^{-2}$ are shown. The green bar depicts the scale $30$~arcsec.} 
\end{figure*}

\addtocounter{figure}{-1}
\begin{figure}
\centering
$\vcenter{\hbox{\includegraphics[width=4cm]{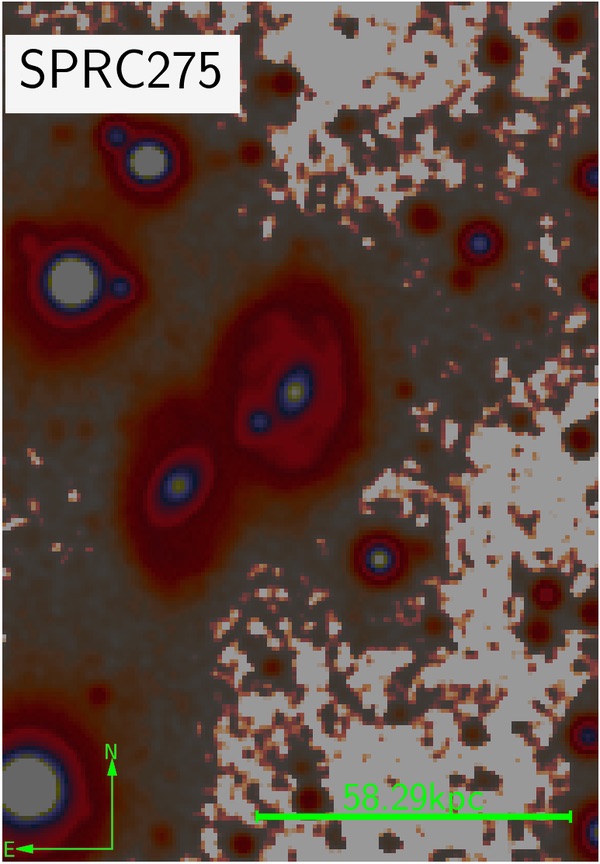}}}$
$\vcenter{\hbox{\includegraphics[width=4cm]{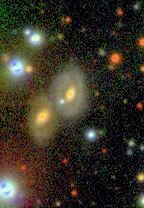}}}$
\caption{(Continued.)}
\end{figure}

In this section, we provide a brief description of each galaxy, with useful information on its structural properties based on our visual inspection of the final stacked images (see Figs.~1 and \ref{fig:legacy_images}) and the results of our multicomponent photometric decomposition which we present below.
To our best knowledge, most galaxies in our sample have little or no information in the literature. The unprecedented depth of the stacked Stripe\,82 images for these galaxies helps us explore very faint features around the selected PRGs in much more detail than previously done. We also compare the Stripe\,82 and Legacy images, in order to confirm the presence of the same LSB features in both surveys. For several galaxies, Subaru Hyper Suprime-Cam \citep{2021arXiv210813045A} observations are available, so we use them as well.

Throughout the paper, we refer to several types of LSB structures, some of which we define here in the following way (see e.g. \citealt{2015MNRAS.446..120D,2019A&A...632A.122M,2020A&A...640A..38R}). Tidal tails are elongated stellar structures which are generated by galaxy interactions and mergers (usually, major mergers). They are generally accepted as evidence for a dynamically cold (and/or often gas-rich disc) component in the accreted companion \citep{1995A&A...297...37C}. Tidal tails are expelled from the primary galaxy following an interaction with a less massive companion. Thus, the material of the tail and the stellar halo of the primary galaxy should have similar properties, such as the same colour. Recently, \citet{2020arXiv200911879R} classified long tidal tails into three shape types: straight, curved and plume (in a form without a certain figure). They inferred that these shapes are associated with different formation mechanisms of tidal tails, e.g. with prograde (curved tails) and retrograde merging (plume tails).

In contrast, tidal streams are related to minor merging events when a low-mass satellite is interacting with a major galaxy and, finally, can be ingested by it. The material of the tidal stellar stream is expelled from the accreted companion and, thus, should have essentially the same colour as the satellite. Also, tidal streams look like long and thin filaments, usually fainter and thinner than tidal tails \citep{2022arXiv220303973S}. However, according to hydrodynamical cosmological simulations in \citet{2019A&A...632A.122M}, the lifetime of streams is, on average, longer (1.5-3\,Gyr) than that of tidal tails, which become visible at a timescale of 0.7-1\,Gyr after the start of the merger.

Also, we use the term `halo', or `envelope' to describe the stellar outskirts of galaxies. This component is mainly a mixture of ancient halo stars formed \textit{in situ} \citep{2011MNRAS.416.2802F,2012MNRAS.420.2245M,2015MNRAS.454.3185C} and accreted stars due to a series of minor merger episodes in the past \citep{2010MNRAS.406..744C,2013MNRAS.434.3348C}. In this paper, we classify outer structures as envelopes if they are remarkably extended and diffuse (see good examples of galaxies with diffuse envelopes in \citealt{2019MNRAS.490.1539R} and \citealt{2020MNRAS.494.1751M}). Among other LSB features which we can notice around and in galaxies are strong asymmetric and disturbed outer structures, arcs, shells, fans, and bridges.

Recently, \citep{2022arXiv220303973S} measured the properties (the area, size, and average surface brightness) for more than 8000 tidal structures around 352 nearby massive galaxies using deep images from the 3.6-meter Canada-France-Hawaii Telescope with approximately the same surface brightness limit as in the SDSS Stripe\,82. One of the important conclusions of their study is that they do not detect any tidal features fainter than 27.5~mag\,arcsec$^{-2}$ in the $r$ band. Therefore, in this study we expect to reveal all such low surface brightness structures around PRGs, if they exist, and describe their properties.

For three PRGs in our sample (SPRC-73, 76 and 238), we quantitatively describe the tidal structures well-seen in the deep images. Here, by $\Psi$ we denote the angle between the major axes of the host and the ring. To estimate some general luminance and geometrical parameters of the stellar structures in these galaxies, we outline them in the stacked images within an isophote of 27 mag\,arcsec$^{-2}$ (we apply blurring with $\sigma=1$~pixel to reduce pixel to pixel variations). We define the average projected width $W_\mathrm{tail}$ and projected length of the tidal structure $L_\mathrm{tail}$ along the centre line, which connects the beginning and the end of the outlined region.  Also, we compute an average surface brightness within the structure and its fraction to the total galaxy luminosity within the same isophote 27.5~mag\,arcsec$^{-2}$. We roughly estimate the colour $g-r$ of the host, ring, and tidal feature within the regions where these structures dominate. The results of our analysis are listed in Table~\ref{tab:streams} and shown in Fig.~A.3 in the Appendix~\ref{Appendix:supplementary}.  As can be seen, the length of the tails of SPRC-73 and SPRC-238 is significantly longer and the width of the tail of SPRC-73 is $\sim4$ times larger than typically found around ETGs (see fig.9 and 10 in \citealt{2022arXiv220303973S}), whereas their average surface brightness is of the same order as measured in the stellar tails and streams.

Finally, we carry out photometric decomposition for the entire sample in the $r$ band. As we show below, using the results of decomposition we can i) analyse the parameters of the PS and the host, ii) reject non-PRGs, and iii) seek the link of these parameters with the origin of the PRGs. Following \citet{rc2015}, we use a S\'ersic function \citep{1963BAAA....6...41S,1968adga.book.....S} to describe both the host and PS.
The free parameters of the S\'ersic function are the surface brightness $\mu_\mathrm{e}$ at the effective (half-light) radius $r_\mathrm{e}$, the S\'ersic index $n$ (which controls the shape of the luminosity profile), the position angle PA, the ellipticity $\epsilon$, and the $C_0$ parameter, which controls the discy/boxy view of the isophote (when $C_0<0$ the isophotes are discy, whereas they look boxy if $C_0>0$ and $C_0=0$ if the isophotes are pure ellipses). The decomposition is done by means of the {\tt{IMFIT}} code \citep{2015ApJ...799..226E} with the full treatment of the extended PSF.

In case when a tidal structure is present, we mask it in our fitting. SPRC-74 and 238 show arc-like features; SPRC-234 demonstrates two faint spiral arm-like structures. Therefore, for these galaxies we mask out these structures in our modelling which implies the presence of a rather symmetric polar component. For SPRC-1 and 275, the overlapping galaxies are fitted as well to better model the light distribution of the target galaxy.
The results of the decomposition for the entire sample are listed in Table~\ref{tab:decomp_all_gals}. In Fig.~\ref{fig:decomp_results} we present our 2D photometric models.

{\bf SPRC-1} 
This is a red early-type galaxy with a polar dust ring which is aligned almost orthogonally to the major axis of the host galaxy ($\Psi=84.5\degr$). Its optical spectrum is typical for active galactic nucleus galaxies with broad emission lines. Its polar dust structure looks like a nearly edge-on dust ring. The visible dust lane noticeably distorts the observed galaxy profile in the central region, therefore we mask it in our modelling. The optical diameter of the ring reaches 18~kpc \citep{2011AstL...37..171R}. SPRC-1 is a member of a compact group of galaxies: it is surrounded by a handful of galaxies which demonstrate apparent tidal features. The outer profile of SPRC-1 is contaminated by the light of the other group members. Therefore, the multiple faint tidal features in the deep image are less apparent than in the other deep images of our sample galaxies. Two spiral galaxies near SPRC-1 show prominent grand-design structures (SDSS\,J000909.89-003705.3 has a strong two-armed spiral, with one arm drawn to its smaller companion, the spiral galaxy SDSS\,J000911.73-003721.8), which might be induced by multiple tidal interactions within this group \citep[see a review by][]{2014PASA...31...35D}. Interestingly, based on their spectra, most large members of this group demonstrate an activity typical for star forming or starburst galaxies. Our decomposition reveals that apart from the polar dust ring, this galaxy also harbours a very luminous polar ($\Psi=82.6\degr$) halo (the polar structure-to-host ratio $L_\mathrm{ps}/L_\mathrm{h}=0.98$) with the S\'ersic index $n=1.7$. This halo may have the same origin as the dust ring since both have very similar position angles. As galaxy interactions and merging are very intensive in compact groups, we can suppose that the formation of the PSs in this S0 galaxy can be related to a wet merger: a head-on collision between two orthogonally oriented galaxies, in which large quantities of gas and dust from another galaxy in this group could produce the polar dust disc and stellar halo we now observe.

{\bf SPRC-4}
It is a kinematically-confirmed PRG \citep{2014ASPC..486...71M}. The deep image of this galaxy reveals the presence of a blue polar-ring which was previously noted by \citet{fink2012}. 
The small S\'ersic index of the PS model $n=0.2$ confirms that this structure is a ring. We notice two chain-like structures around SPRC-4 which seem to be spiralling towards the host galaxy. In fact, these structures can be parts of fainter extended loops of gaseous and stellar material.  
\citet{fink2012} note at least 10 galaxies in the neighbourhood of SPRC-4, some with the signatures of ongoing star formation. 
A close spiral galaxy to SPRC-4, SDSS\,J024256.42-005658.4, has, in fact, a different redshift (0.181) and thus cannot be gravitationally bound to SPRC-4.
The nearest blue dwarf galaxy (its morphological type is not clear due to poor resolution), SDSS\,J024252.07-005646.6, is located at a projected distance of about 80~kpc and can be a potential donor of the gas for SPRC-4.
 We suggest that the ongoing star formation in the ring and its formation itself can be explained by the (tidal) accretion of gas or disruption of a gas-rich dwarf. In the case of tidal accretion, we suspect the blue dwarf galaxy SDSS\,J024252.07-005646.6 to be the donor (although its redshift is unknown).
 
{\bf SPRC-69 (II Zw 92) } 
Its deep image displays that the outer, almost polar wide ring ($\Psi=77.5\degr$, $n=0.3$) surrounds the central body ($n=2.4$). The results of spectral observations confirm the presence of a PS, in both ionised gas and stars \citep{moiseev2011}. Along the major axis of the central body, \citet{moiseev2011} also point out a linear increase in the rotation velocity of stars. Our deep and colour images clearly show faint bound structures (spirals?) which start from the ends of the polar-ring. This may indicate that the polar ring is in fact a disc. There are two close dwarf galaxies (SDSS\,J204807.12+000416.3 and SDSS\,J204806.63+000332.8), but we cannot see any faint structural features related to these objects. Also, their spectroscopic redshifts are unknown. The regular structure of SPRC-69 down to very low surface brightnesses provides proof that this system is relatively old and relaxed. We suppose that the polar ring in this galaxy formed due to either a major merger (the contribution of the ring is substantial $L_\mathrm{ps}/L_\mathrm{h}=0.60$) or a tidal accretion followed by a merger with the donor after the formation of the ring \citep{rs1997,bourcomb2003}. The survival time for tidal tails and streams is up to $\approx2-3$~Gyr \citep{2019A&A...632A.122M,2020A&A...640A..38R}. Apparently, SPRC-69 should have started its formation earlier, so that we do not observe any fine structures around this galaxy in the current epoch.

{\bf SPRC-73} 
The spectrum of SPRC-73 is typical for star forming galaxies.
The major axis of the diffuse PS is inclined by $\Psi\sim63\degr$ with respect to the major axis of the host galaxy. 
There is also a bright blue point source (a bulge or a nucleus of the victim galaxy?) to the north of the centre which lies exactly in the plane of the diffuse polar component and has the same redshift $z=0.059$ as the host galaxy. 
In addition, we can see a faint arc with another red dwarf galaxy SDSS\,J003208.55+010834.5 to the west. Many small extended objects around SPRC-73 may also be involved in the multiple (wet and dry) minor merging with the host galaxy (for example, there is a faint bridge with SDSS\,J003207.63+010847.4 at the north-western outskirts of SPRC-73).  
This galaxy shows a giant curved tail in the deep image which bends around the host and extends about one arcmin ($\sim70$~kpc) to the south (see also \citealt{fink2012}). The average surface brightness of the curved tail $\langle \mu_\mathrm{r} \rangle\sim26.6$~mag\,arcsec$^{-2}$ is beyond the depth of single-exposure SDSS frames, and, thus, it cannot be seen in them. 
We note that the colours of the polar ring and the tail are very similar and differ from the reddish colour of the host. Thus, this tail is intimately related to the observed PS (its ``current'' profile is purely exponential $n=1.0$). 
Therefore, we can conclude that the most plausible scenario to explain the forming PS in SPRC-73 is ongoing merging with a galaxy, the debris of which has not yet been dispersed. We find that the luminosity of the PS, including the tail and the nucleus, is rather high as compared to the host ($L_\mathrm{ps+}/L_\mathrm{h}=0.75$). This is an indication of the major merger which leads to the formation of the PS in SPRC-73.

{\bf SPRC-74}
A blue asymmetric elongated structure crosses the main body of SPRC-74 at an angle of about 75$\degr$ to the major axis. The PS is strongly bent and displaced relative to the centre of the host galaxy, which shows a S\'ersic index close to exponential ($n=1.3$). We can distinguish a diffuse, LSB galaxy of irregular shape on the north-eastern side of SPRC-74 and a faint short stream to the east. The main body of the host galaxy demonstrates distorted outer isophotes and bends towards the diffuse LSB galaxy in the north-east. The disturbed shape of both the main and diffuse galaxies hints at the ongoing gravitational interaction. As noted in \citet{fink2012}, a close companion SDSS\,J004805.83-001251.2 lies at a projected distance of about 1.6 arcmin ($\sim$100~kpc) from SPRC-74. In our stacked image we can see an extended, thin stream directed to the west, which connects multiple clumpy objects (tidal dwarf galaxies?) up to 1~arcmin from the galaxy. Unfortunately, beyond that radius the scattered light from a close bright star pollutes our SDSS image. However, the Subaru Hyper Suprime-Cam image in Fig.~A.2 shows much less scattered light from the bright star, so we can distinguish a faint ($\langle \mu_\mathrm{r} \rangle\sim26.6$~mag\,arcsec$^{-2}$) thin bridge connecting SPRC-74 and the closest blue spiral galaxy SDSS\,J004805.83-001251.2 ($z=0.05661$). This fact leads us to suggest that the most probable scenario for this PRG is tidal accretion from the gas-rich neighbour.

\begin{table}
\bc
\begin{minipage}[]{0.86\textwidth}
\caption[]{Properties of the tidal tails for the three polar-ring galaxies measured using the SDSS $r$-band images.\label{tab:streams}}\end{minipage}
\setlength{\tabcolsep}{1pt}
\small
  \begin{tabular}{ccccccccccc}
  \hline\noalign{\smallskip}
   Galaxy  & $\Psi$ & $L_\mathrm{tail}$ & $L_\mathrm{tail}$& $W_\mathrm{tail}$& $W_\mathrm{tail}$ &  $\langle \mu_\mathrm{r,tail} \rangle$ & $f_\mathrm{tail}$ & $(g-r)_\mathrm{tail}$ & $(g-r)_\mathrm{h}$ & $(g-r)_\mathrm{ps}$  \\ 
           &    (deg)&   (arcsec)  & (kpc) & (arcsec) & (kpc) &  (mag\,arcsec$^{-2}$) & & (mag) & (mag) & (mag)                \\
    (1)    & (2)         & (3)   & (4)   &  (5)     & (6) & (7) & (8) & (9) & (10) & (11)   \\
\hline\noalign{\smallskip}
SPRC-73    & 63 & 63.7 & 72.8 & 32.3 & 36.6 & $26.5\pm0.3$ & 0.032 & $0.61\pm0.22$ & $0.82$ & $0.64$\tabularnewline
SPRC-76    & 86 & 29.0 & 24.9 & 11.7 & 10.0 &  $25.6\pm0.7$ & 0.089 & $0.41\pm0.12$ & $0.99$ & $0.48$\tabularnewline
SPRC-238   & 82 & 37.1 & 43.1 & 9.2 & 10.7 & $26.3\pm0.3$ & 0.018 & $0.69\pm0.20$ & $0.60$ & $0.38$\tabularnewline
  \noalign{\smallskip}\hline
  \end{tabular}
  \ec
  \tablecomments{0.86\textwidth}{     (1) Galaxy name;
     (2) angle between the major axes of the polar ring and host;
     (3), (4) projected length of the centreline of the tail;
     (5), (6) average projected width of the tail;
     (7) average surface brightness within the tail; 
     (8) tail-to-total luminosity (within the isophote 27 mag\,arcsec$^{-2}$)  ratio;
     (9), (10), (11) - colours of the tail, host and polar ring, respectively.}
 \end{table} 

{\bf SPRC-76}
The red host galaxy in SPRC-76 is surrounded by the blue material of the merging galaxy,
the debris (its profile has $n=1.1$) of which we observe as the blue object SDSS\,J012129.26+003732.6 (it has a concordant $z=0.04386$) with the blue curved tail emerging from it. Also, we can see a clear bridge (depicted by a maroon colour in Fig.~1) which connects the main body and the red satellite SDSS\,J012129.86+003719.7. Furthermore, we detect a plume on the eastern side of SPRC-76 to the north, up to another dwarf galaxy SDSS\,J012130.98+003745.6. Possibly, there is also a link with the spiral galaxy SDSS\,J012130.98+003745.6 to the north-east. We also detect a faint bridge with the LSB galaxy SDSS\,J012132.24+003758.6 which is located beyond the previously mentioned spiral in the same direction. The strongly distorted structure of SPRC-76 indicates that we observe a major merger which is likely to be responsible for the ongoing generation of the PS.

{\bf SPRC-77}
\citet{fink2012} note on SPRC-77 and 185 that they ``resemble `normal' disc galaxies with a symmetric bright disc around their central bulge rather than typical PRGs''. However, careful examination of the deep image and its photometric decomposition reveals that SPRC-77 has a distinctive oval central component (the host), which is embedded in the blue wide polar ring ($n=0.1$).
This galaxy demonstrates quite regular inner and outer isophotes. 
Our deep image also reveals an oval envelope ($n=1.1$), the major axis of which seems to be misaligned with the major axis of the ring\footnote{We will consider this galaxy, along with SPRC-185, in a separate paper. Here we only present the results of our multicomponent (host+ring+halo) decomposition.}.
We can see a small south-eastern arc in the plane of the disc which can be a dwarf satellite stretched by the tidal forces. 
The profiles of SPRC-77 and two south-western interacting galaxies seem to have either a bridge or overlapping, so there is not enough evidence to suggest that this system is gravitationally bound. 
Also, we observe several dwarf galaxies on the north-eastern side of SPRC-77 which exhibit signatures (tidal streams and bridges) of interaction with the major galaxy. The regular shape of this envelope signifies that it has the same origin as the polar ring -- a major merger in the past.

{\bf SPRC-185}
SPRC-185 is another pure edge-on galaxy in our sample. The smooth outer structure, which is almost orthogonal to the central galaxy, becomes prominent only in our deep image. It has a boxy shape and does not resemble ``normal'' polar ring galaxies  from our sample. Our decomposition model for this galaxy consists of a nucleus, a dominant bulge, a flared disc, and a polar component with the S\'ersic index $n\approx 1.5$. Thus, the PS in this galaxy is not a ring, but a halo. We also note a plume on the eastern side of the galaxy, which is probably related to the formation of the polar halo due to a merger.

{\bf SPRC-186}
Our decomposition shows that this galaxy has a reddish inner disc (a pseudobulge with $n\approx1$), a blue inner ring and two spiral arms (as a single component with $n=0.6$), instead of an outer ring which might be interpreted as tilted with respect to the inner component. Therefore, this suggests that SPRC-186 is not a PRG. Unlike most polar-ring galaxies in our sample, we see no LSB details around it.

{\bf SPRC-188}
This highly inclined galaxy with a prominent dust lane shows a polar edge-on structure ($\Psi\sim83\degr$), which is slightly off the centre and is more pronounced in the south direction than to the north. We also see several dwarf galaxies which show bridges with SPRC-188. The galaxy SDSS\,J033405.00+010541.0 to the west in the plane of the main body of SPRC-188 demonstrates a tidal feature which goes outwards. The overall shape is highly asymmetric and disturbed, possibly due to the recent formation of the PS (the ring also looks disturbed but does not have any fine structures connected to it) and interaction with several large neighbours (the irregular galaxy SDSS\,J033406.02+010615.0 to the north and the triplet of interacting galaxies to the east -- all of them have concordant redshifts $z\sim0.048$). We also note the prominent envelope around SPRC-188. Nevertheless, the overall profile of the host is exponential ($n=1.0$) and red (probably due to dust, but the resolution is not enough to see a prominent dust lane). The profile of the PS is gaussian ($n=0.5$) and quite blue ($g-r=0.22$). We suggest that the PS in SPRC-188 formed due to an encounter of the host with a less massive spiral galaxy ($L_\mathrm{ps}/L_\mathrm{h}=0.22$). This event occurred quite long ago ($\gtrsim2-3$~Gyr), so that we do not observe prominent tidal structures related to the polar ring.

{\bf SPRC-234}
The central object in SPRC-234 is likely a highly inclined S0 galaxy ($n=1.8$). The ordinary SDSS image of this galaxy does show a tilted structure with respect to the central body. However, the deep image displays it as a very faint two-armed spiral rather than a ring. Nevertheless, there is also a possibility that these spiral arms are tightly wound into a ring, similar to NGC\,660. The average surface brightness along these arms is indeed very low ($\langle \mu_{r} \rangle \sim26$~mag\,arcsec$^{-2}$ and the estimated arms-to-total ratio is unusually low: $f_{r}\sim0.07$). This is inconsistent with what we observe as a rule in grand-design galaxies: \citet{2020MNRAS.493..390S} find $\langle \mu_{r} \rangle \ll 25$~mag\,arcsec$^{-2}$ and $f=0.21\pm0.07$. The arms seem rather diffuse with no bright clumps of star formation. If we suppose that the red central object is not an S0 galaxy but a bar, its shape looks very unusual: the isophotes are pure elliptical in the inner region and become discy at the periphery, whereas bars in galaxies usually look boxy, but exceptions are also found \citep[see e.g.][and references therein]{2020arXiv200709090S}. Furthermore, galaxy bars usually demonstrate truncated profiles along the major axis and their S\'ersic indices are $n=0.8\pm0.2$ \citep{2009MNRAS.393.1531G}. Therefore, we can make a conclusion that either this object is a PRG host or that we observe a random projection of a lenticular galaxy on an open spiral. 

We can see several small galaxies in the vicinity of SPRC-234 but their redshifts are unknown and we do not see any signs of interaction with the host galaxy. We can conclude that the nature of this object is uncertain and special observations are required.

{\bf SPRC-238} 
This galaxy shows multiple signatures of interaction with its neighbours which distorted its shape, such as an arc above the galaxy plane in the north-west direction. Perhaps, the other side of this PS is covered by the galaxy disc, so that we cannot observe it. The deep image demonstrates that there is a pronounced straight tidal tail to the east of the galaxy with a north-east direction. The colour of this tail is similar to that of the host galaxy. 
The most plausible scenario for generating the observed PS in SPRC-238 is tidal disruption of a gas-rich dwarf ($L_\mathrm{ps}/L_\mathrm{h}\lesssim0.1$), the debris of which is either covered by the host galaxy body, or completely stretched by the tidal forces, so that we do not see any nucleus left (although we note a small clumpy structure right above the plane).

{\bf SPRC-275}
This galaxy is visually connected to its close twin, SDSS\,J205354.20-005814.3, although we cannot determine its redshift and confirm their real gravitational bound. Also, we do not see any faint structures in their vicinity. Although these two galaxies look similar, the main difference between them is the bluish colour of the ring ($n=0.2$) in SPRC-275, which points to a different origin of the observed structure. The central component in this galaxy is poorly resolved but its S\'ersic index $n=1.5$ is not typical of a bar. However, we can see a nucleus which may be, in fact, a tiny bulge embedded in a bar. The ring-to-host ratio $L_\mathrm{ps}/L_\mathrm{h}=1.5$ is extremely large -- only few galaxies out of the 50 PRGs in \citet{rc2015} have a ratio as large as 1.3. Also, the angle between the major axes of the host and the ring is the smallest in the sample $\Psi=19\degr$ \citep[see also][]{2013AstBu..68..371S}. Therefore, we surmise that the blue ring in this galaxy lies in reality in the plane of the central structure which consists of a tiny bulge and a bar.

\begin{table}
\bc
\begin{minipage}[]{0.86\textwidth}
\caption[]{The results of our decomposition for the Stripe\,82 images in the $r$ passband. For SPRC-77, we only provide the total magnitude of the host since it is not resolved. In SPRC-185, the host is an S0 galaxy, which consists of a nucleus, a bulge, and a disc, and, therefore, we only present the total magnitude for the superposition of these components. For SPRC-74, SPRC-234, and SPRC-238, the fitting was only carried out for the central galaxy. The total magnitudes and colours have been corrected for Galactic extinction and $k$-correction.\label{tab:decomp_all_gals}}\end{minipage}
\setlength{\tabcolsep}{1pt}
\small
\begin{tabular}{ccccccccccc} 
 \hline\noalign{\smallskip}
SPRC &  $m_\mathrm{h}$ & $r_\mathrm{e,h}$ & $n_\mathrm{h}$ & $(g-r)_\mathrm{h}$ & $m_\mathrm{ps}$ & $r_\mathrm{e,ps}$ & $n_\mathrm{ps}$ & $(g-r)_\mathrm{ps}$ & $\Psi$ & PS/Host   \\ 
     &  (mag)          &    (arcsec)     &                &                    &   (mag)         &   (arcsec)       &  & &   (deg)  &  \\
  (1)     & (2)    & (3) & (4) & (5) & (6) & (7) & (8) & (9) & (10) & (11)  \\ 
\hline\noalign{\smallskip}
1 & 17.21 & 1.2 & 2.4 & 0.80 & 16.13 & 12.8 & 1.7 & 0.76 & 82.6 & 0.98\\
4 & 16.18 & 1.9 & 2.1 & 0.65 & 16.27 & 2.7 & 0.2 & 0.29 & 80.4 & 0.17\\
69 & 15.59 & 1.3 & 2.4 & 0.77 & 16.14 & 14.2 & 0.3 & 0.57 & 77.5 & 0.60\\
73 & 16.26 & 1.9 & 2.8 & 0.82 & 16.78 & 7.8 & 1.0 & 0.64 & 63.3 & 0.62\\
74 & 16.20 & 1.7 & 1.3 & 0.73 & --- & --- & --- & --- & $75^{*}$ & ---\\
76 & 17.38 & 1.4 & 6.1 & 0.99 & 17.82 & 4.2 & 0.7 & 0.48 & 86.0 & 0.67\\
77 & 17.93 & --- & --- & 1.27 & 17.53 & 2.9 & 0.1 & 0.47 & 89.2 & 0.39\\
185 & 15.84 & --- & --- & 0.74 & 17.58 & 4.1 & 1.5 & 0.66 & 70.6 & 0.20\\
186 & 16.09 & 1.3 & 1.0 & 0.51 & 15.34 & 3.6 & 0.6 & 0.82 & 27.1 & 2.00\\
188 & 15.57 & 6.7 & 1.0 & 1.01 & 17.21 & 10.0 & 0.5 & 0.22 & 83.2 & 0.22\\
234 & 16.75 & 2.1 & 1.8 & 0.72 & --- & --- & --- & --- & --- & ---\\
238 & 15.59 & 3.3 & 1.5 & 0.54 & --- & --- & --- & --- & --- & ---\\
275 & 18.41 & 1.1 & 1.5 & 0.86 & 17.68 & 6.7 & 0.2 & 0.31 & 18.9 & 1.45\\
\noalign{\smallskip}\hline
\end{tabular}
\ec
\tablecomments{0.86\textwidth}{
   (1) Galaxy name, 
   (2) total $r$-band magnitude of the host, 
   (3) effective radius of the host, 
   (4) S\'ersic index of the host,
   (5) $g-r$ colour of the host,
   (6) total $r$-band magnitude of the PS, 
   (7) effective radius of the PS, 
   (8) S\'ersic index of the PS,
   (9) $g-r$ colour of the PS, 
   (10) angle between the major axes of the PS and the host (for SPRC-74 it is estimated by hand), 
   (11) the PS-to-host luminosity ratio.}
\end{table}

\section{Discussion}
\label{sec:discussion}

In Sect.~\ref{sec:results}, we provided a detailed description of each galaxy in our sample. Exploring the properties of the host and the PS using photometric decomposition of the $r$-band Stripe\,82 images allows us to draw some important conclusions.

The results for our sample are generally compatible with those obtained for a different sample of 50 PRGs by \citet{rc2015} and 31 PRGs by \citet{rm2019} who used the same approach of modelling PRGs with two pure S\'ersic components for a host and a ring\footnote{Note that not all galaxies from the mentioned studies have estimates of the parameters for the host and PSs, as in some cases the host is too small to be fitted (only total magnitude is provided) or the polar ring is too faint to be modelled. In total, we collected the S\'ersic parameters for 70 hosts and 40 PSs.}. The ratio of the luminosity of the PS to the luminosity of the host varies in a wide range from 0.17 (SPRC-4) to 0.97 (SPRC-1), with an average value $0.50\pm0.27$ ($0.38\pm0.31$ for the \citealt{rc2015} sample), excluding SPRC-186 and 275. Our photometric decomposition of the latter two galaxies suggests that due to the outlying values of some of their parameters (e.g. the low projected angle between the major axes of the host and the PS and the very high host-to-ring ratio) these galaxies cannot be considered PRGs. The hosts in our sample are mostly lenticular galaxies with $n=1.9\pm0.6$ ($n=2.3\pm1.0$ in \citealt{rc2015}).

One of the main contributions of this paper is an analysis of the main parameters of the S\'ersic model for the PSs. The S\'ersic index is an important indicator of the morphology of a galaxy structure, which, for PSs, helps us conclude whether the luminous matter is distributed in the form of a ring ($n\lesssim0.5-0.7$) or it has some other geometry. In Fig.~2 we show the distribution by the S\'ersic index of the polar components for the galaxies from our sample and those from \citet{rc2015} and \citet{rm2019}. As one can see, there are two groupings of the PSs: true polar rings with $n\lesssim0.5-0.7$ and PSs with $n\gtrsim1$, which may be polar discs (SPRC\,28, see Fig.~3), polar spheroids (SPRC\,1 and SPRC\,185), and only forming PSs (SPRC\,73). Indeed, a photometrically distinctive polar ``ring'' can appear a kinematically cold component, a disc, as shown by \citet{2002AJ....123..195I} for NGC\,4650A. SPRC\,185, as discussed in Sect.~\ref{sec:results}, has a red faint envelope which is significantly tilted with respect to the S0 host galaxy.

\begin{figure}
\label{fig:sersic_index}
\centering
\includegraphics[width=8.5cm]{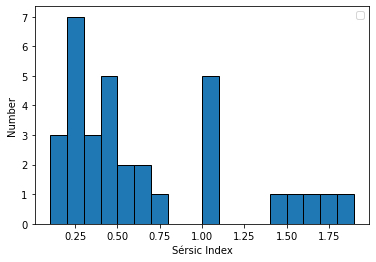}
\caption{The distribution by the S\'ersic index of PSs for the joined samples of polar-ring galaxies from this study, \citet{rc2015}, and \citet{rm2019}.}
\end{figure}

\begin{figure}
\label{fig:sprc28}
\centering
\includegraphics[width=8cm]{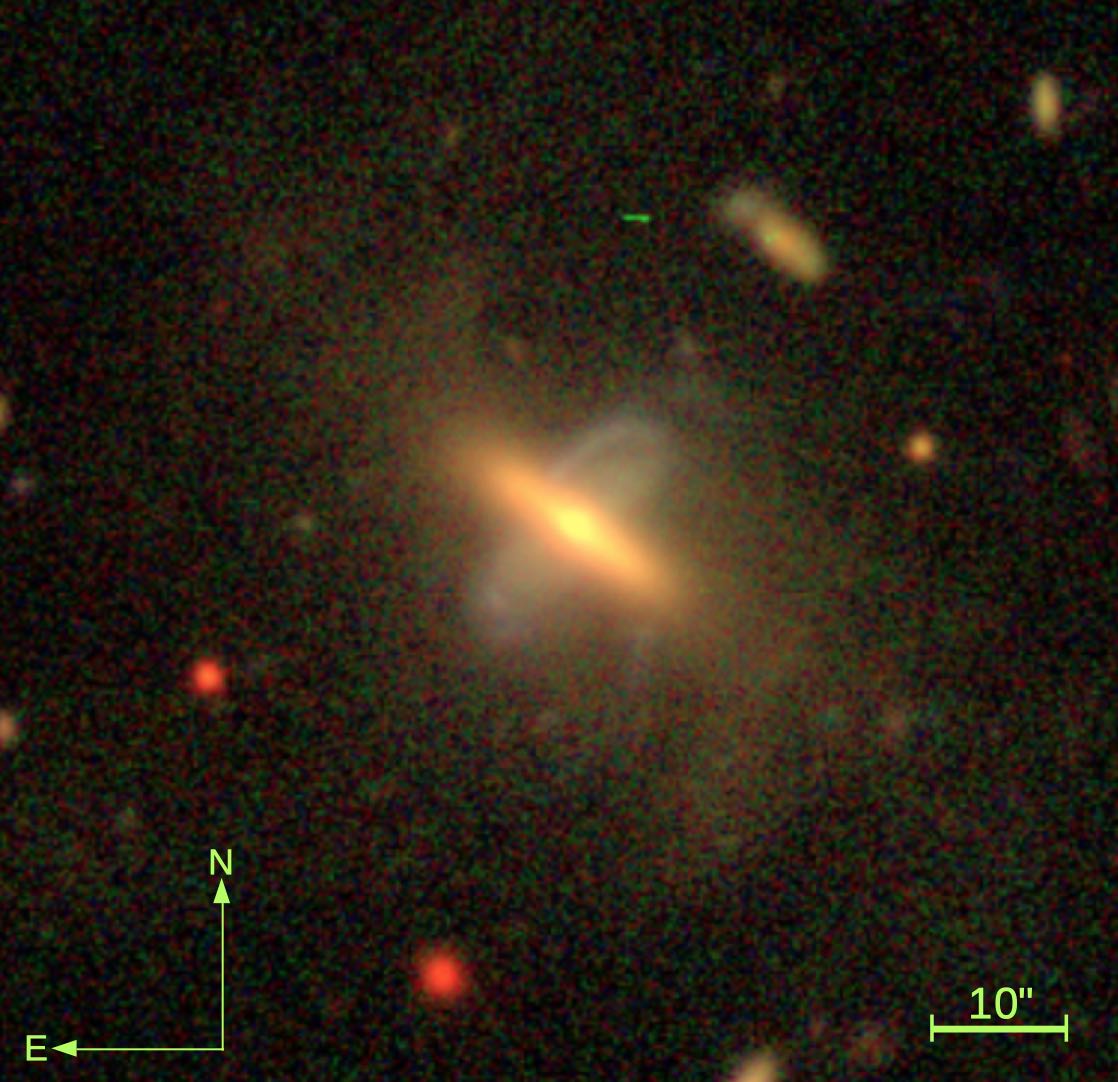}
\caption{An RGB image for SPRC\,28 from Subaru Hyper Suprime-Cam \citep{2021arXiv210813045A}. Note that the host galaxy has a faint disturbed envelope which consists of multiple tidal debris of past merger event(s).}
\end{figure}

Although we know some PRGs in which the host and the ring have similar masses (for example, SPRC-3 and SPRC-35), usually the luminosity of the ring is 2-3 mag fainter than that of the host galaxy \citep[see e.g.][]{fink2012,rc2015}. However, our sample comprises rather bright polar rings as compared to their hosts: polar rings are mostly 0.5-1 mag fainter. Therefore, as our results suffer from the obvious selection bias, we limit ourselves to consider only those mechanisms which seem appropriate to drive the evolution of the sample galaxies.

In Sect.~\ref{sec:results}, we quantitatively described some general properties of the well-visible tidal features around three PRGs in our sample. Despite the large extent of these structures (up to $\sim70$~kpc for SPRC-73), they typically contain only about 5\% of the total galaxy light. This is consistent with what is usually observed around early-type galaxies \citep{2005AJ....130.2647V}. The LSB structures in and around the sample galaxies, also noted by us in Sect.~\ref{sec:results}, can provide important clues to the formation of these galaxies.
In the Introduction, we have summarised the main scenarios to form PRGs. Below we briefly discuss these mechanisms in the context of our results.


{\bf Major merging.} 
In our sample, six galaxies (SPRC-1, 69, 73, 76, 77 and 185) show some evidence for this scenario. SPRC-1 resides in a very conducive environment for galaxy-galaxy interactions and mergers, in a compact group \citep{1997ARA&A..35..357H}. Obviously, this galaxy has undergone one or several mergers. As we have shown in Sect.~\ref{sec:results}, one of the most remarkable results which we obtained in our modelling, is that its polar dust ring is aligned with the polar stellar halo. We argue in favour of the scenario in which this PRG formed due to a major merger when one of the merging components was a gas- and dust-rich galaxy, whereas the second progenitor was a usual early-type galaxy. SPRC-73 and 76 are in process of the formation of their PSs, and we can see a nucleus of the progenitor in this major merger. However, the question of whether these structures will be stable and not dissolve with time, creating an elliptical galaxy, remains open. In the case of SPRC-69, which has settled into an equilibrium, the polar ring might have formed due to a major merger, as the mass of the progenitor for the polar ring (possibly polar disc?) is considerable. An alternative scenario for this galaxy may be gas accretion.

{\bf Tidal accretion.}
This mechanism is the most plausible explanation for SPRC-74 and 238. These galaxies show very faint bridges of the material falling from the donor onto the PRG, similar to what is observed in the pair NGC\,3808 \citep{ord2016}. This mechanism finds good observational support in the literature \citep[see e.g.][]{rbc2006,2008AstBu..63..201M,2012MNRAS.427.2772F} and, thus, may explain some other observed PRGs in our sample (SPRC-4 and 69) where the polar rings are in a quasi-stable state, but no apparent accretion is seen in our deep images. 

{\bf Cold accretion.} 
In this scenario, PRGs form through the accretion of external cold gas from cosmic web filaments \citep[see e.g.][]{2007MNRAS.382.1809B,stan2009,2010ApJ...714.1081S}. Gas infall along a filament may have a narrow angular momentum distribution and will settle into a ring structure \citep{brook2008}. In principle, this mechanism can explain the formation of SPRC-4, 69 and 234 (if the latter galaxy is a true PRG). However, only deep photometric data are not sufficient to examine this scenario.

{\bf Disruption of a satellite.}
This scenario can be applicable to the explanation of the observed blue polar ring in SPRC-4 (taking into account the observed faint spiral trails around this galaxy) and the discy PS in SPRC-188. Also, one or several minor merger events can produce a halo around a galaxy \citep[see e.g.][]{2010MNRAS.406..744C}. We propose that tidal disruption of accreted dwarf galaxies coupled with the misalignment of the triaxial (oblate) dark matter halo \citep{2002ApJ...576...21V,2003ApJ...597...35C,2005ApJ...628...21S,2012ApJ...750..107S} and its inner stellar disc might lead to the formation of a tilted stellar halo in this galaxy (see the discussion in \citealt{2020MNRAS.497.2039M} and references therein).

\section{Summary}
\label{sec:summary}

We have analysed deep observations for 13 candidates to polar-ring galaxies by exploiting deep images from SDSS Stripe\,82 and other deep surveys. We summarise our results as follows:
\begin{enumerate}
\item For the selected galaxies, we enhanced the Stripe\,82 data by stacking the three $gri$ bands together. This allowed us to reach the depth $\mu_\mathrm{r}\approx30$\,mag\,arcsec$^{-2}$.
\item All galaxies under study, except SPRC-69, 186 and 275, demonstrate low surface brightness structures in or around them: tidal features in a group (SPRC-1), tidal tails and streams (SPRC-4, 73, 74, 76, 185 and 238), arcs (SPRC-74 and 238), faint envelopes (SPRC-28, 77 and 185), extremely faint spiral arms or a (un)closed ring (SPRC-234). We suspect that LSB structures are probably nearly ubiquitous in these rare objects.
\item For SPRC-73, 76 and 188 we can directly observe the formation of the polar ring due to major or minor merging (see also Fig.~4, right plot). In SPRC-73 and 76 we observe a possible debris of the victim galaxy. We provide a qualitative description (see Sect.~\ref{sec:results}) and a quantitative analysis for three galaxies with distinctive tidal tails -- SPRC-73, 76 and 238 (see Table~\ref{tab:streams}).
\item In some cases (e.g. SPRC-69 and SDSSJ\,015828.87+013549.9 in Fig.~4, left plot), there are no visible tidal features even in ultra-deep images, but the faint haloes and disrupted envelopes (see also Fig.~3) suggest that the merger occurred several Gyr ago.
\item SPRC-74 demonstrates very faint signs of tidal accretion (see Fig.~A.2). This is an additional proof of the importance of this mechanism in an effort to explain the formation of PRGs.
\item We carried out detail photometric decomposition for all sample galaxies in the $r$ passband to simultaneously fit the host and the PS. The advantage of this study is that we extract from the literature and study the structural parameters for both the polar rings and the hosts (see Table~\ref{tab:decomp_all_gals}). This helps us estimate not only the morphology of the host, but also the morphology of the ring. 10 out of 40 galaxies from this study, \citet{rc2015}, and \citet{rm2019} harbour PSs with a S\'ersic index $n\geq1$ (see Fig.~2). These PSs can be either spheroids or discs. Some of these structures are still forming: in many of them we can discern a disrupted core of the merging galaxy.
\end{enumerate}

In our follow-up study, we are about to expand the sample of PRGs for which deep observations have been obtained.
With upcoming large area sky surveys such as the Vera C. Rubin Observatory \citep[formerly the Large Synoptic Survey Telescope,][]{2019ApJ...873..111I}, astronomy will soon be faced with a tremendous wealth of deep observations, which should help to reveal more LSB features around PRGs and test the existing formation mechanisms of these unique objects.

\begin{figure}
\label{fig:add_galaxies}
\centering
\includegraphics[width=8.5cm]{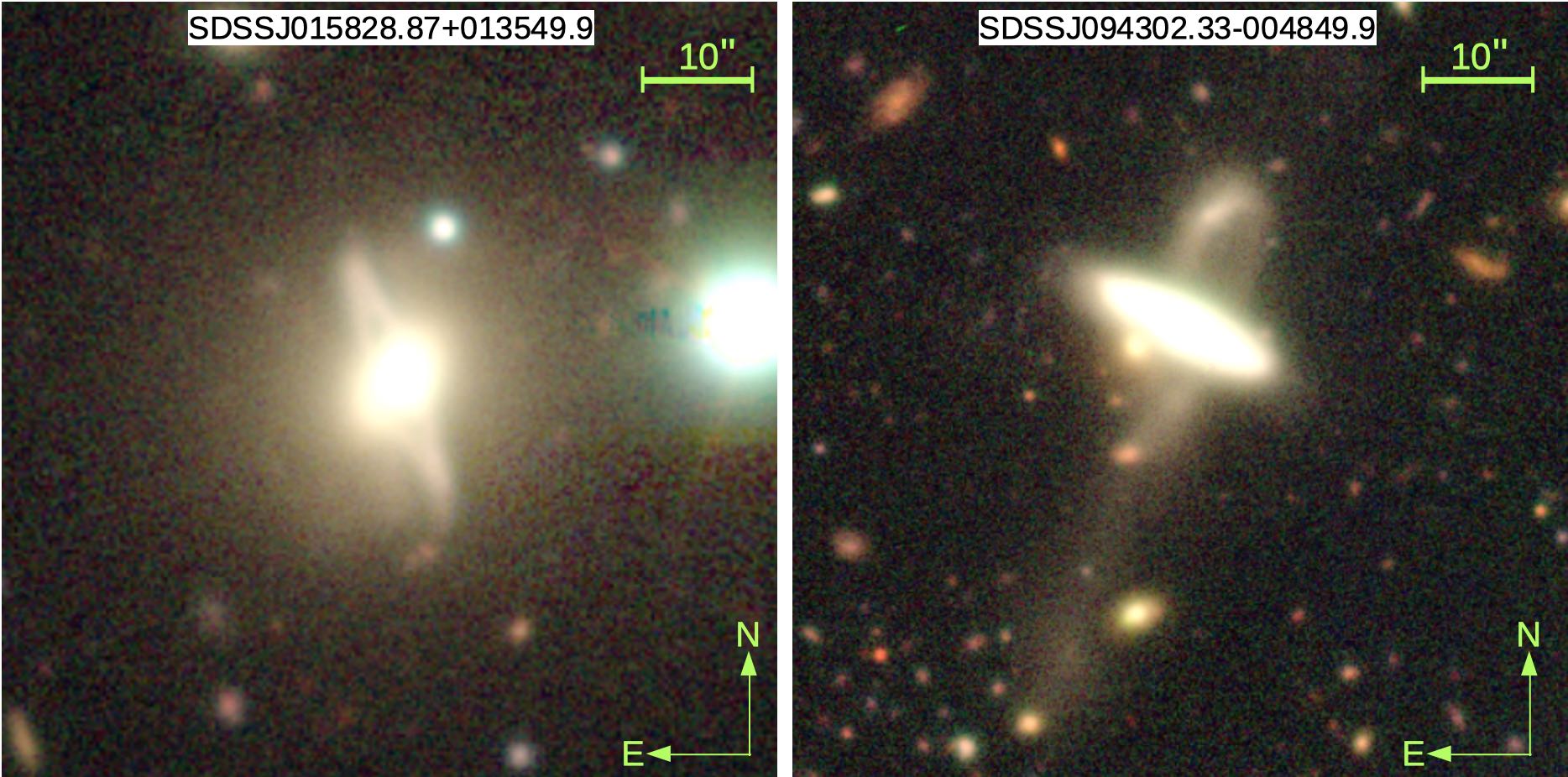}
\caption{RGB images for two PRGs from \citet{rm2019} based on  Subaru Hyper Suprime-Cam observations. The left-hand image shows an envelope around the host and a warp of the polar ring. The right-hand image demonstrates the formation of a PS via a minor merger.}
\end{figure}

\begin{acknowledgements}

Aleksandr Mosenkov and Vladimir Reshetnikov acknowledge financial support from the Russian Science Foundation
(grant no. 22-22-00483) for developing an algorithm of enhancing deep optical imaging using modern sky surveys.

This research has made use of the NASA/IPAC Infrared Science Archive (IRSA; \url{http://irsa.ipac.caltech.edu/frontpage/}), and the NASA/IPAC Extragalactic Database (NED; \url{https://ned.ipac.caltech.edu/}), both of which are operated by the Jet Propulsion Laboratory, California Institute of Technology, under contract with the National Aeronautics and Space Administration.  This research has made use of the HyperLEDA database (\url{http://leda.univ-lyon1.fr/}; \citealp{2014A&A...570A..13M}). 
This work is based in part on observations made with the {\it Spitzer} Space Telescope, which is operated by the Jet Propulsion Laboratory, California Institute of Technology under a contract with NASA. 

Funding for the Sloan Digital Sky Survey IV has been provided by the Alfred P. Sloan Foundation, the U.S. Department of Energy Office of Science, and the Participating Institutions. SDSS-IV acknowledges
support and resources from the Center for High-Performance Computing at
the University of Utah. The SDSS web site is www.sdss.org.

SDSS-IV is managed by the Astrophysical Research Consortium for the 
Participating Institutions of the SDSS Collaboration including the 
Brazilian Participation Group, the Carnegie Institution for Science, 
Carnegie Mellon University, the Chilean Participation Group, the French Participation Group, Harvard-Smithsonian Center for Astrophysics, 
Instituto de Astrof\'isica de Canarias, The Johns Hopkins University, Kavli Institute for the Physics and Mathematics of the Universe (IPMU) / 
University of Tokyo, the Korean Participation Group, Lawrence Berkeley National Laboratory, 
Leibniz Institut f\"ur Astrophysik Potsdam (AIP),  
Max-Planck-Institut f\"ur Astronomie (MPIA Heidelberg), 
Max-Planck-Institut f\"ur Astrophysik (MPA Garching), 
Max-Planck-Institut f\"ur Extraterrestrische Physik (MPE), 
National Astronomical Observatories of China, New Mexico State University, 
New York University, University of Notre Dame, 
Observat\'ario Nacional / MCTI, The Ohio State University, 
Pennsylvania State University, Shanghai Astronomical Observatory, 
United Kingdom Participation Group,
Universidad Nacional Aut\'onoma de M\'exico, University of Arizona, 
University of Colorado Boulder, University of Oxford, University of Portsmouth, 
University of Utah, University of Virginia, University of Washington, University of Wisconsin, 
Vanderbilt University, and Yale University.

The Legacy Surveys consist of three individual and complementary projects: the Dark Energy Camera Legacy Survey (DECaLS; NOAO Proposal ID \# 2014B-0404; PIs: David Schlegel and Arjun Dey), the Beijing-Arizona Sky Survey (BASS; NOAO Proposal ID \# 2015A-0801; PIs: Zhou Xu and Xiaohui Fan), and the Mayall z-band Legacy Survey (MzLS; NOAO Proposal ID \# 2016A-0453; PI: Arjun Dey). DECaLS, BASS and MzLS together include data obtained, respectively, at the Blanco telescope, Cerro Tololo Inter-American Observatory, National Optical Astronomy Observatory (NOAO); the Bok telescope, Steward Observatory, University of Arizona; and the Mayall telescope, Kitt Peak National Observatory, NOAO. The Legacy Surveys project is honored to be permitted to conduct astronomical research on Iolkam Du'ag (Kitt Peak), a mountain with particular significance to the Tohono O'odham Nation.

NOAO is operated by the Association of Universities for Research in Astronomy (AURA) under a cooperative agreement with the National Science Foundation.

This project used data obtained with the Dark Energy Camera (DECam), which was constructed by the Dark Energy Survey (DES) collaboration. Funding for the DES Projects has been provided by the U.S. Department of Energy, the U.S. National Science Foundation, the Ministry of Science and Education of Spain, the Science and Technology Facilities Council of the United Kingdom, the Higher Education Funding Council for England, the National Center for Supercomputing Applications at the University of Illinois at Urbana-Champaign, the Kavli Institute of Cosmological Physics at the University of Chicago, Center for Cosmology and Astro-Particle Physics at the Ohio State University, the Mitchell Institute for Fundamental Physics and Astronomy at Texas A\&M University, Financiadora de Estudos e Projetos, Fundacao Carlos Chagas Filho de Amparo, Financiadora de Estudos e Projetos, Fundacao Carlos Chagas Filho de Amparo a Pesquisa do Estado do Rio de Janeiro, Conselho Nacional de Desenvolvimento Cientifico e Tecnologico and the Ministerio da Ciencia, Tecnologia e Inovacao, the Deutsche Forschungsgemeinschaft and the Collaborating Institutions in the Dark Energy Survey. The Collaborating Institutions are Argonne National Laboratory, the University of California at Santa Cruz, the University of Cambridge, Centro de Investigaciones Energeticas, Medioambientales y Tecnologicas-Madrid, the University of Chicago, University College London, the DES-Brazil Consortium, the University of Edinburgh, the Eidgenossische Technische Hochschule (ETH) Zurich, Fermi National Accelerator Laboratory, the University of Illinois at Urbana-Champaign, the Institut de Ciencies de l'Espai (IEEC/CSIC), the Institut de Fisica d'Altes Energies, Lawrence Berkeley National Laboratory, the Ludwig-Maximilians Universitat Munchen and the associated Excellence Cluster Universe, the University of Michigan, the National Optical Astronomy Observatory, the University of Nottingham, the Ohio State University, the University of Pennsylvania, the University of Portsmouth, SLAC National Accelerator Laboratory, Stanford University, the University of Sussex, and Texas A\&M University.

BASS is a key project of the Telescope Access Program (TAP), which has been funded by the National Astronomical Observatories of China, the Chinese Academy of Sciences (the Strategic Priority Research Program "The Emergence of Cosmological Structures" Grant \# XDB09000000), and the Special Fund for Astronomy from the Ministry of Finance. The BASS is also supported by the External Cooperation Program of Chinese Academy of Sciences (Grant \# 114A11KYSB20160057), and Chinese National Natural Science Foundation (Grant \# 11433005).

The Legacy Survey team makes use of data products from the Near-Earth Object Wide-field Infrared Survey Explorer (NEOWISE), which is a project of the Jet Propulsion Laboratory/California Institute of Technology. NEOWISE is funded by the National Aeronautics and Space Administration.

The Legacy Surveys imaging of the DESI footprint is supported by the Director, Office of Science, Office of High Energy Physics of the U.S. Department of Energy under Contract No. DE-AC02-05CH1123, by the National Energy Research Scientific Computing Center, a DOE Office of Science User Facility under the same contract; and by the U.S. National Science Foundation, Division of Astronomical Sciences under Contract No. AST-0950945 to NOAO.

The Hyper Suprime-Cam (HSC) collaboration includes the astronomical communities of Japan and Taiwan, and Princeton University. The HSC instrumentation and software were developed by the National Astronomical Observatory of Japan (NAOJ), the Kavli Institute for the Physics and Mathematics of the Universe (Kavli IPMU), the University of Tokyo, the High Energy Accelerator Research Organization (KEK), the Academia Sinica Institute for Astronomy and Astrophysics in Taiwan (ASIAA), and Princeton University. Funding was contributed by the FIRST program from the Japanese Cabinet Office, the Ministry of Education, Culture, Sports, Science and Technology (MEXT), the Japan Society for the Promotion of Science (JSPS), Japan Science and Technology Agency (JST), the Toray Science Foundation, NAOJ, Kavli IPMU, KEK, ASIAA, and Princeton University. 

This paper makes use of software developed for Vera C. Rubin Observatory. We thank the Rubin Observatory for making their code available as free software at http://pipelines.lsst.io/.

This paper is based on data collected at the Subaru Telescope and retrieved from the HSC data archive system, which is operated by the Subaru Telescope and Astronomy Data Center (ADC) at NAOJ. Data analysis was in part carried out with the cooperation of Center for Computational Astrophysics (CfCA), NAOJ. We are honored and grateful for the opportunity of observing the Universe from Maunakea, which has the cultural, historical and natural significance in Hawaii. 

\end{acknowledgements}

\bibliographystyle{raa}
\bibliography{article}

\appendix
\section{Supplementary images}
\label{Appendix:supplementary}

\begin{figure*}
\centering
\includegraphics[width=12cm]{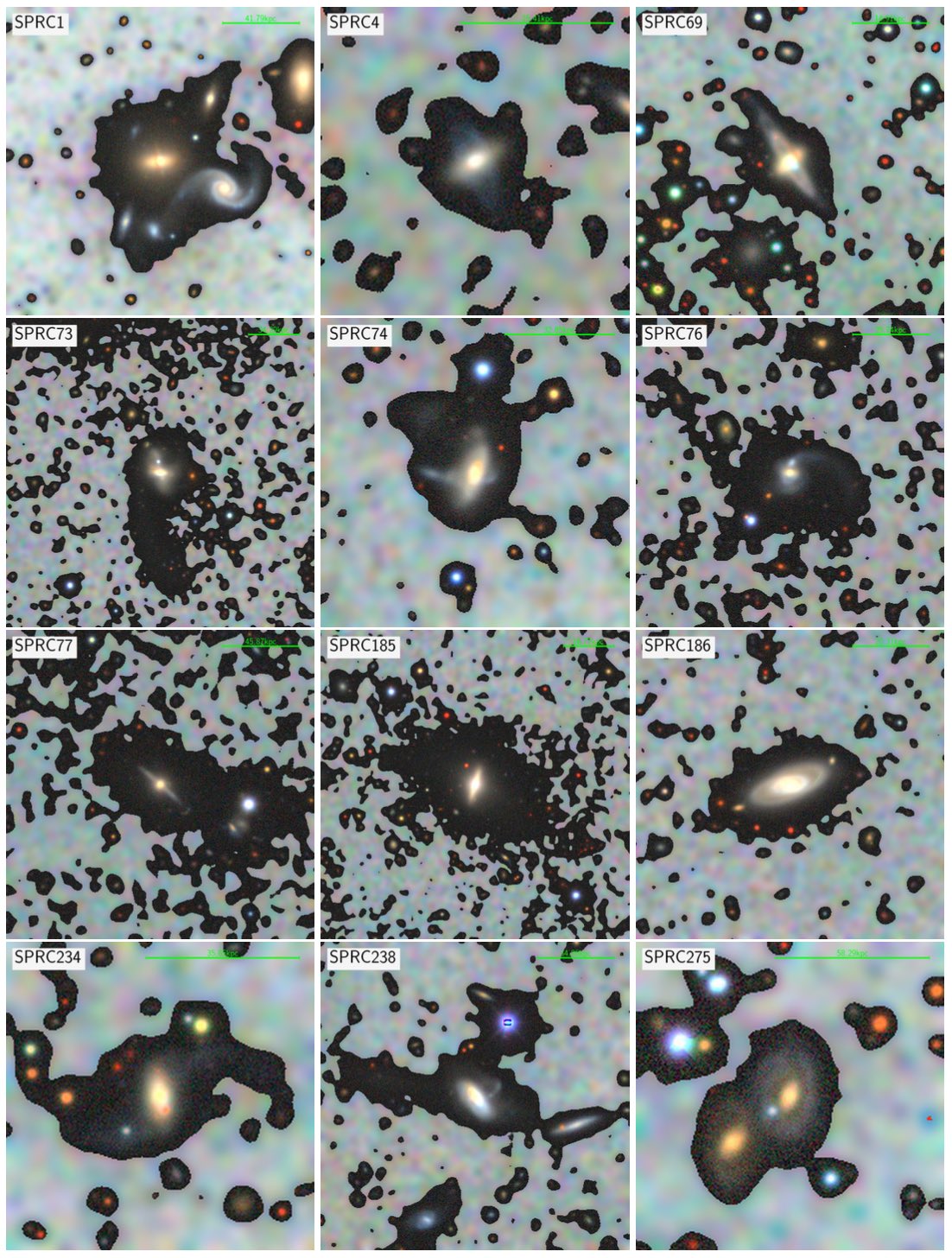}
\caption{Enhanced DESI Legacy images for the PRG sample from this study. The darkish shadows correspond to surface brightnesses brighter than 27~mag\,arcsec$^{-2}$ in the $r$ band. North is up and east is to the left. The scale bar depicts 30~arcsec.} \label{fig:legacy_images}
\end{figure*}

\begin{figure*}
\label{fig:ssp_pic}
\centering
$\vcenter{\hbox{\includegraphics[width=14cm]{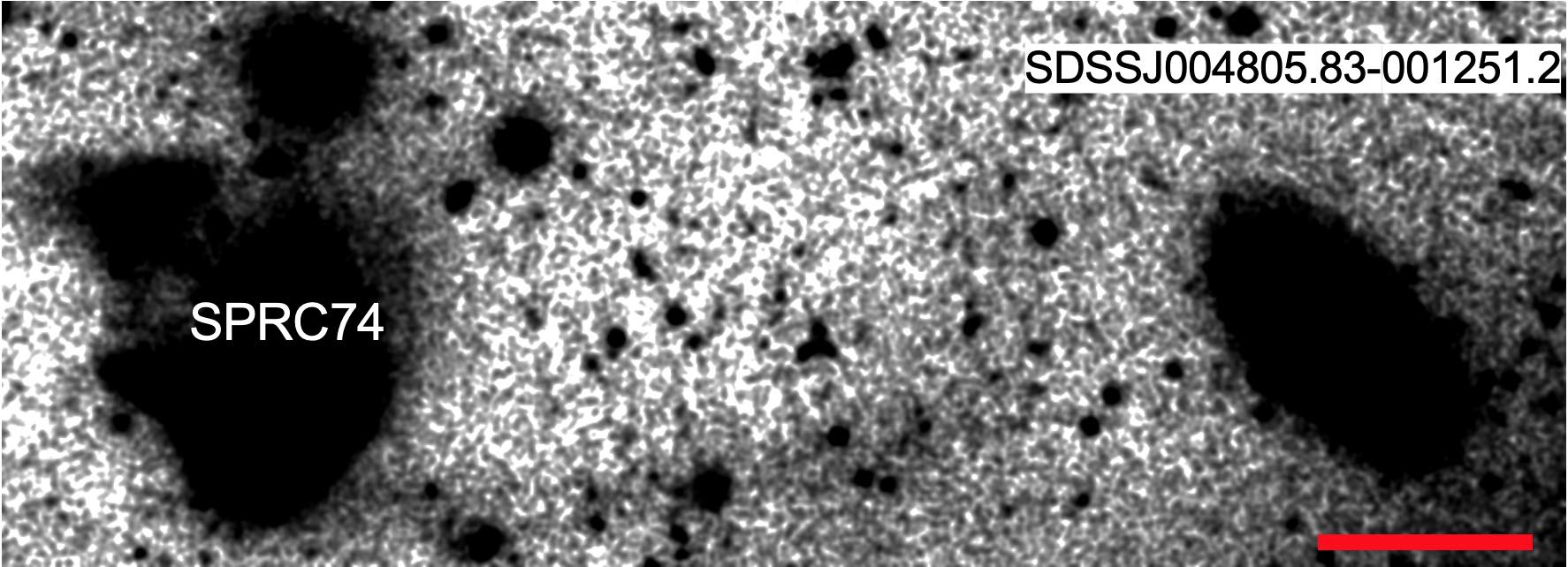}}}$
\caption{An RGB image of SPRC-74 from Subaru Hyper Suprime-Cam. North is up and east is to the left. The length of the bar in the bottom-right corner is 20~arcsec.}  
\end{figure*}

\begin{figure*}
\label{fig:stellar_streams}
\centering
$\vcenter{\hbox{\includegraphics[width=14cm]{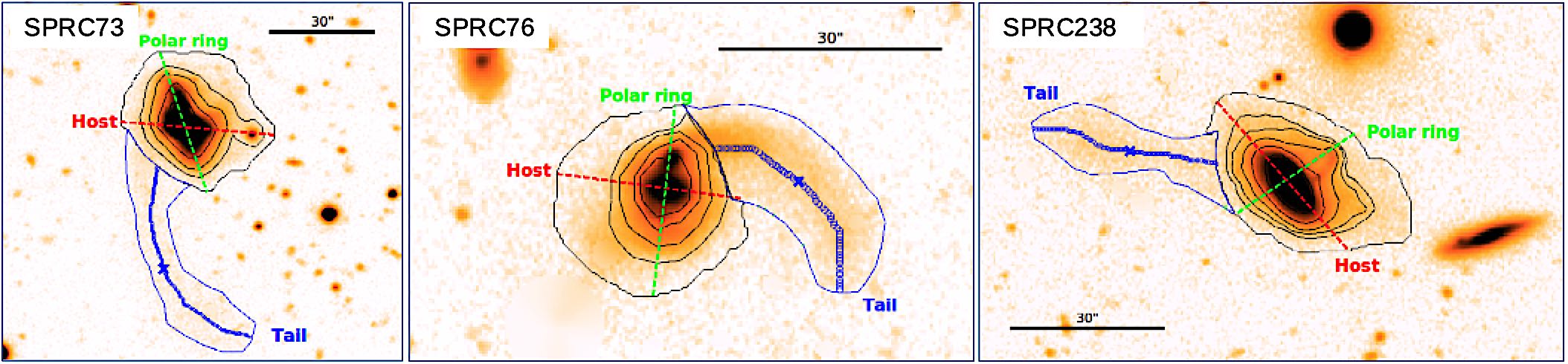}}}$
\caption{Description of the tidal tails in SPRC-73, SPRC-76, and SPRC-238. The dashed lines depict the axes of the structures: red -- for the host, green -- for the polar ring. The blue little circles show the centreline of the tail in each image. North is up and east is to the left.}  
\end{figure*}

\begin{figure*}
\centering
$\vcenter{\hbox{\includegraphics[width=17cm]{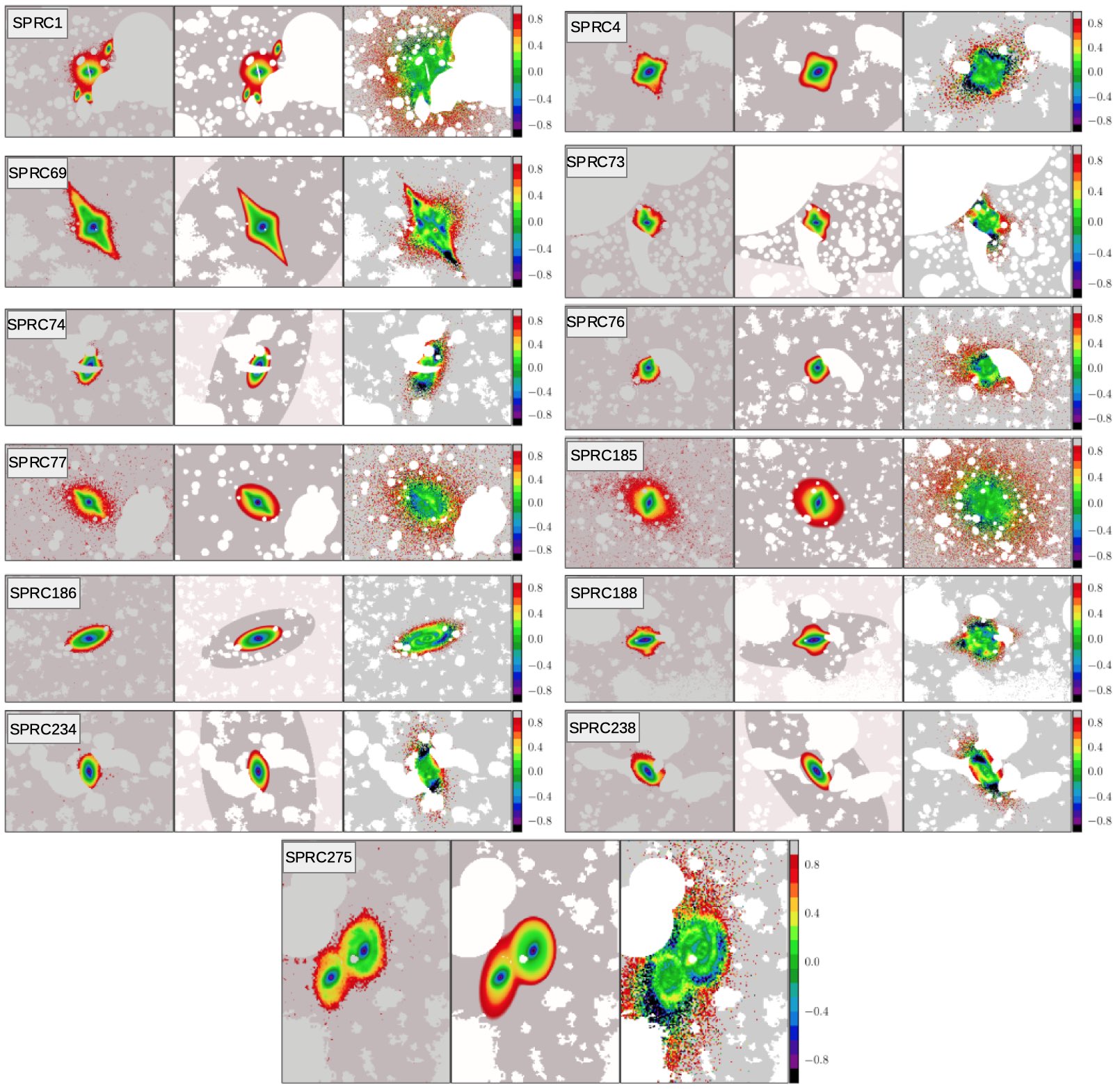}}}$
\caption{Results of our photometric decomposition for all sample galaxies in the $r$ band: the observation (left), the model (middle), and the residual image (right), which indicates the relative deviation between the fit and the image.} \label{fig:decomp_results}
\end{figure*}

\label{lastpage}

\end{document}